\documentclass[sigconf]{acmart}

\usepackage{booktabs}
\usepackage{algorithm}
\usepackage{algpseudocode}
\usepackage{bbold}
\usepackage{diagbox}
\usepackage{graphicx}
\usepackage{subcaption}
\usepackage{enumitem}
\usepackage{amssymb}

\newcommand{\todo}[1]{\textcolor{red}{#1}}

\newcommand{\sz}[1]{\textcolor{blue}{#1}}

\usepackage{xspace}
\newcommand{\empt}{\texttt{Empty}\xspace}

\usepackage{enumitem}
\setlist[itemize,1]{leftmargin=0.4cm}

\newtheoremstyle{mydef}
{2ex}
{2ex}
{\itshape}
{}
{\scshape}
{: }
{0.5em}
{}
\theoremstyle{mydef}

\begin{document}

\copyrightyear{2019} 
\acmYear{2019} 
\acmConference[CIKM '19]{The 28th ACM International Conference on Information and Knowledge Management}{November 3--7, 2019}{Beijing, China}
\acmBooktitle{The 28th ACM International Conference on Information and Knowledge Management (CIKM '19), November 3--7, 2019, Beijing, China}
\acmPrice{15.00}
\acmDOI{10.1145/3357384.3357932}
\acmISBN{978-1-4503-6976-3/19/11}

\fancyhead{}
\title{Auto-completion for Data Cells in Relational Tables}

\author{Shuo Zhang}
\affiliation{%
  \institution{University of Stavanger}
}
\email{shuo.zhang@uis.no}

\author{Krisztian Balog}
\affiliation{%
  \institution{University of Stavanger}
}
\email{krisztian.balog@uis.no}

\begin{abstract}

We address the task of auto-completing data cells in relational tables.  Such tables describe entities (in rows) with their attributes (in columns).  We present the \textsc{CellAutoComplete} framework to tackle several novel aspects of this problem, including: (i) enabling a cell to have multiple, possibly conflicting values, (ii) supplementing the predicted values with supporting evidence, (iii) combining evidence from multiple sources, and (iv) handling the case where a cell should be left empty.  Our framework makes use of a large table corpus and a knowledge base as data sources, and consists of preprocessing, candidate value finding, and value ranking components.  Using a purpose-built test collection, we show that our approach is 40\% more effective than the best baseline.  

\end{abstract}



\maketitle

\section{Introduction}
\label{sec:int}

\begin{figure}[t]
\vspace*{0.5\baselineskip}
   \centering
   \includegraphics[width=0.48\textwidth]{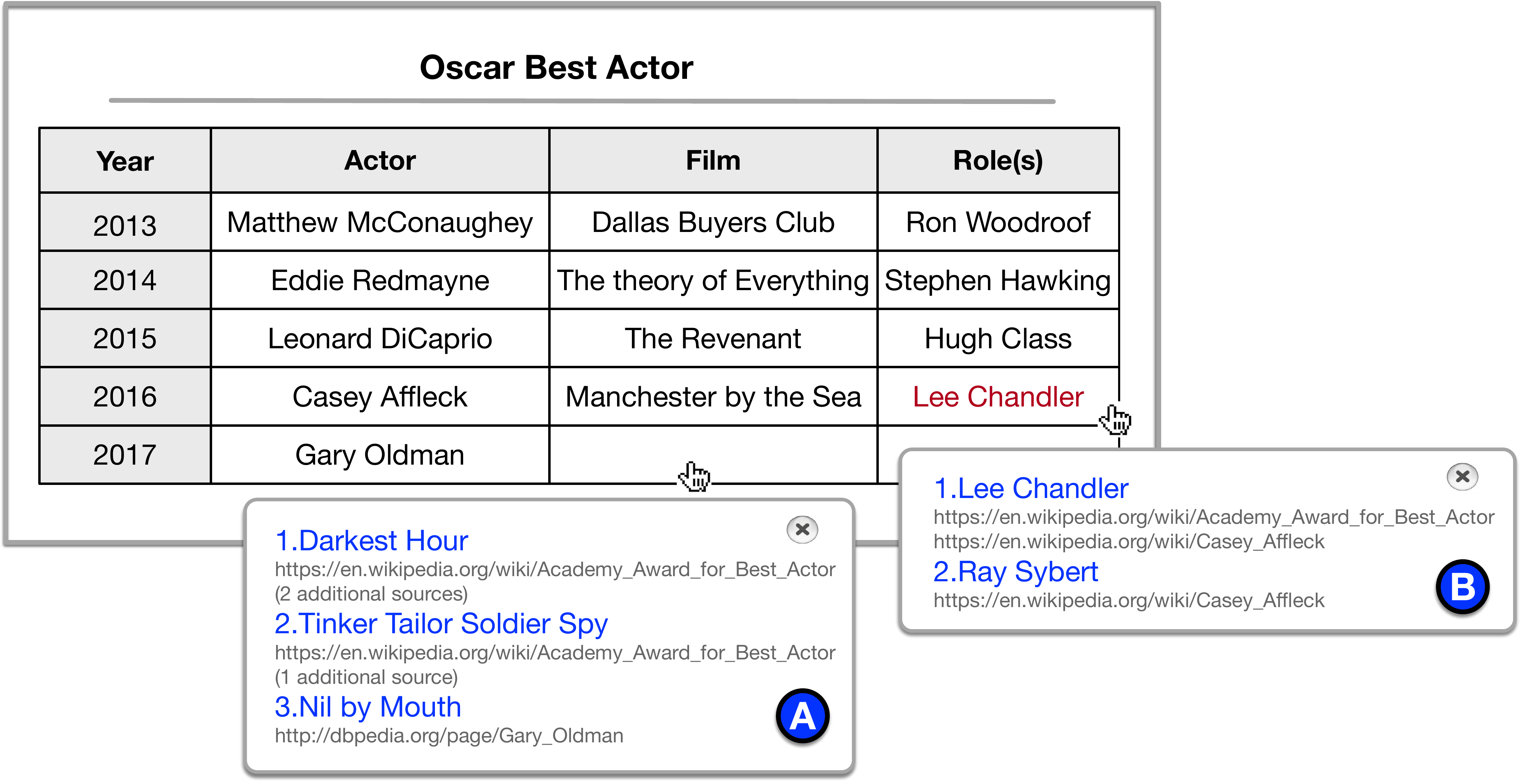} 
   \vspace*{-1.5\baselineskip}
   \caption{Envisioned user interface. By clicking on a table cell, the user receives a ranked list of suggested values along with supporting sources. Case (A) is to find the value for an empty cell. Case (B) is to check/verify an existing value.}
   \label{fig:ui}
   \vspace*{-1\baselineskip}
\end{figure}

Tables are a frequently used tool for collecting information about entities of interest.  
Relational tables (also referred to as entity-attribute tables~\citep{Yakout:2012:IEA}) are a particular type of table utilized for that purpose, where rows correspond to entities and columns corresponds to attributes of those entities. 
There is a growing body of research on assisting users in the labor-intensive process of table creation by helping them
to augment tables with data~\citep{Zhang:2017:ESA, Yakout:2012:IEA, Ahmadov:2015:THI},
retrieve existing tables~\citep{Zhang:2018:AHT, Ahmadov:2015:THI, Yakout:2012:IEA}, 
and even automatically generate entire tables~\citep{Zhang:2018:OTG}.
This paper falls in the category of \emph{data augmentation} (also referred to as \emph{data imputation}~\citep{Ahmadov:2015:THI}), which is concerned with extending a given input table with more data. 
Specifically, we propose to equip spreadsheet applications with a kind of \emph{auto-complete} functionality, where, upon clicking on a cell, the user is presented with a list of suggestions of possible values for that cell; see Fig.~\ref{fig:ui}.
This task has in fact been addressed in~\citep{Zhang:2017:ESA} for row and column heading cells of relational tables.
Our goal in this paper is to provide a similar auto-complete feature for \emph{value cells}. This is a considerably more difficult problem that poses a range of unique challenges.

First, it is paramount that each value that is offered to the user in the list of suggestions for a given cell has \emph{supporting evidence}.  That is, the user can verify that value by tracing it back to its source of origin.  Intuitively, the more support a given value has the higher it should be ranked on the list of suggestions.
This renders model-based approaches, which train a machine learning model to guess the missing values~\citep{Ahmadov:2015:THI}, unsuitable for this scenario.
Second, it is important to recognize when a cell should in fact be left empty, and to not deceive the user with nonsense or misleading suggestions. 



%
%

%
%

Addressing these challenges requires a fundamentally different approach from prior work.
Our \textsc{CellAutoComplete} framework consists of two main steps.  First, we identify candidate values from a table corpus and from a knowledge base. 
A key component in this process is mapping the target attribute (e.g., ``venue'') to (i) column heading labels in the table corpus that have the same meaning (e.g., ``stadium'') and (ii) to predicates in a knowledge base (e.g., \texttt{<dbo:ground>}).
Second, we combine numerous signals in a learning-to-rank (LTR) framework to generate a ranking of the candidate values.
A particularly novel idea, that is captured by a specific group of features, is to consider the semantic similarity of each candidate table, which mentions the target entity and attribute, to the input table.
In order to deal with cells that should be left empty, we introduce a special designated \empt value.  This allows us to quantify our belief that a given cell should be left empty.
Our experimental results show that our LTR approach outperforms the best single-source baseline by about 40\% in terms of NDCG@10.


%
%

%
\begin{figure*}[t]
\vspace*{-1.5\baselineskip}
   \centering
   \includegraphics[width=0.85\textwidth]{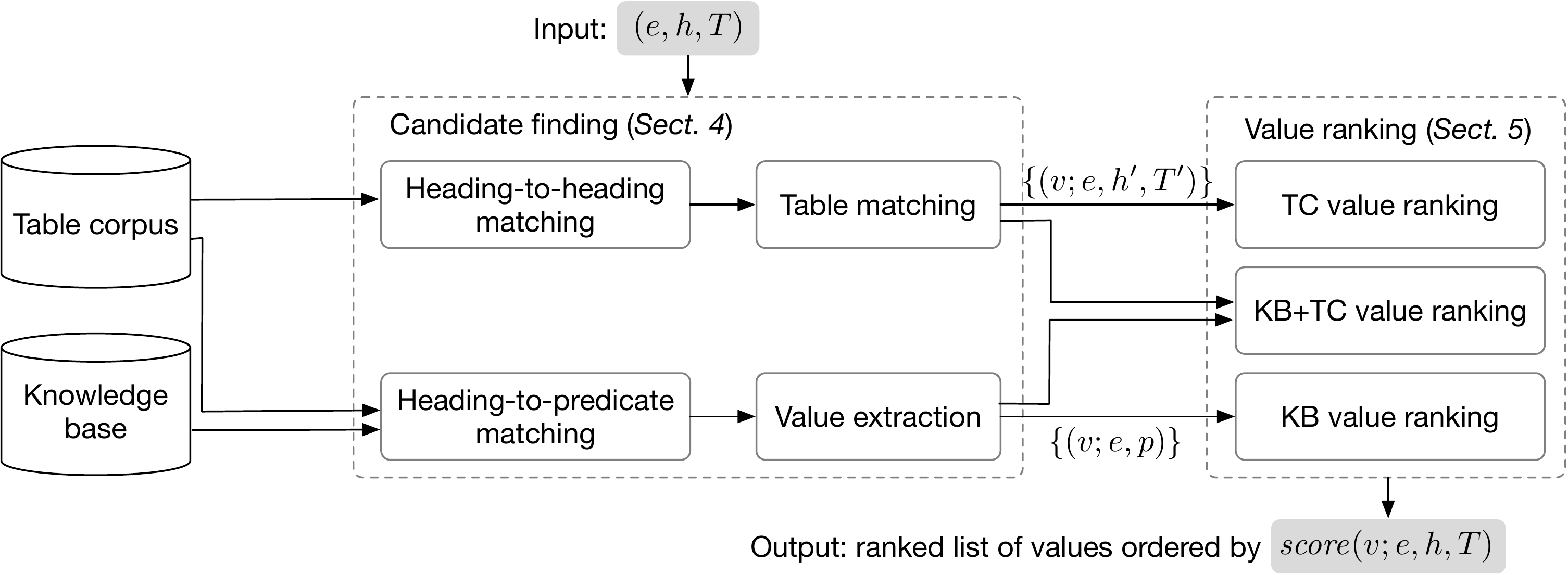} 
   \vspace*{-0.6\baselineskip}
   \caption{Overview of the \textsc{CellAutoComplete} framework.}
   \label{fig:pipeline}
   \vspace*{-0.3\baselineskip}
\end{figure*}

Our contributions can be summarized as follows:
\begin{itemize}
	\item We present the \textsc{CellAutoComplete} framework for finding cell values in relational tables (Sect.~\ref{sec:ps}), which consists of  candidate value finding (Sect.~\ref{sec:cf}) and value ranking (Sect.~\ref{sec:vr}) components.  Specific novel technical contributions include the heading-to-heading and heading-to-predicate matching components (\S\ref{sec:cf:h2h} and \S\ref{sec:cf:h2p}) as well as the features designed for combining evidence from multiple sources and for predicting empty values (Sect.~\ref{sec:vr-comb}).
	\item We develop a purpose-built test collection based on Wikipedia tables, comprising 35k manually labeled cell-value pairs for 1000 cells (Sect.~\ref{sec:expsetup}), and perform an extensive experimental evaluation (Sect.~\ref{sec:eval}).  Our experiments show that our LTR approach substantially outperforms existing data augmentation techniques.
\end{itemize}
All resources developed in this study are made publicly available at \url{https://github.com/iai-group/cikm2019-table}.

\vspace*{-0.5\baselineskip}
\section{Related Work}
\label{sec:rw}


\emph{Table augmentation} is the task of extending an input table with more data~\citep{Zhang:2017:ESA,Venetis:2011:RST,Yakout:2012:IEA,Zhang:2013:ISM, Cafarella:2008:WEP, DasSarma:2012:FRT, Lehmberg:2015:MSJ, Bhagavatula:2013:MEM, Ahmadov:2015:THI}. 
This task may be performed on different levels: (i) simply finding related tables, (ii) augmenting individual cells, such as populating heading rows and columns with additional entities and column labels or finding values for data cells, and (iii) adding entire rows/columns.

To find related tables that are potentially useful, \citet{DasSarma:2012:FRT} search for \emph{entity complement} tables that are semantically related to entities in the input table, and \emph{schema complement} tables for augmenting table schema.  
Aiming for more targeted (cell-level) augmentation, \citet{Zhang:2017:ESA} propose intelligent assistance functionalities for populating row and column headings of relational tables. Specifically, they develop probabilistic models for ranking a list of suggestions, that is, entities to be added as headings of new rows and labels to be added as headings of new columns.  Both subtasks rely on a knowledge base and on similar tables retrieved from a table corpus as sources of suggestions.
A recent study utilizes Word2vec to train embeddings for these tasks~\citep{Deng:2019:TNW}.
\citet{Yakout:2012:IEA} present the InfoGather system that performs table augmentation in three flavors: \emph{augmentation by example}, \emph{schema auto-complete} and \emph{augmenting attributes}, designed for augmenting table entities, heading labels, and table cell values, respectively. 
The former two tasks are the same as those in~\citep{Zhang:2017:ESA}.
The last of these, augmenting attributes, is particularly relevant for the current paper.  The approach taken in~\citep{Yakout:2012:IEA} is to first search for similar tables, then to fuse the corresponding cell values from the matching tables. 
\citet{Yakout:2012:IEA} match entities by value overlap. 
To match column labels, they propose a holistic approach of utilizing additional information including similarities based on context, attribute names, and column values, to overcome the shortcomings of traditional schema matching and instance level features.
We use InfoGather  as a baseline in our experiments.
With a similar mission, \citet{Zhang:2013:ISM} extend the system as InfoGather+, focusing on numerical and time-varying attributes augmentation.

The above tasks are extending tables cell by cell. There is also a body of work on joining (entire) rows/columns from existing tables.
\citet{Bhagavatula:2013:MEM} propose the \emph{relevant join} task, which returns a ranked list of column triplets for a given input table. The first two elements are the input column and the matched column, and the third element quantifies the correlation between them. The remaining columns that co-exist with the matched columns are taken as candidates.
They employ a linear model to rank the candidate columns, which could be joined to the input table as an additional column. \citet{Lehmberg:2015:MSJ} design a join search engine, which searches related tables based on table column headings, then applies a series of left outer joins by taking columns from returned tables and adding them to the input table.
Focusing on table cell values, \citet{Ahmadov:2015:THI} propose a combined method, depending on the features of missing values, so as to look missing values from web data, predict them utilizing machine learning models, or combine the above two to find the most likely values.

While it is not a table augmentation approach, the work by~\citet{Zhang:2018:OTG} is also highly relevant to our task.  The authors propose to answer entity-oriented queries by generating relational tables as answers ``on the fly.'' They first determine the entities (heading rows) and their attributes (heading columns), then find the values of the corresponding cells. They build a value catalog based on matching values found in Wikipedia tables and in DBpedia, giving priority to the latter, from which they can then fetch cell values. We use their method as a baseline in our experiments.

Our work is also related to \emph{truth/fact finding} research.
\citet{Yin:2011:FFL} build a fact lookup engine, FACTOR, based on web tables. FACTOR can answer fact lookup queries, which are about a certain attribute value of one entity, e.g., the birth date of Taylor Swift. Specifically, they extract entity-attribute-value triples from tables on the Web, aggregate and clean them, and store them in a database.  For a given query, FACTOR generates every possible combination of entity and attribute, and retrieves a set of values, which are ranked considering entity similarity, attribute similarity, URL format, and value types.
\citet{Ernst:2018:HHF} present HighLife to harvest higher-arity facts from texts, in order to capture more complete and deeper knowledge about events or multi-entity relationships. The method is distantly supervised by seed facts, which are facts that have been verified.  
Facts found in tables can be used for question answering. For example,~\citet{Sun:2016:TCS} propose a deep matching model for matching question and table cells. The matched table cells are taken as the answers.
These approaches take free text queries as input, while our task considers a table as input. This makes the nature of the task quite different, and renders those approaches unsuitable for us as baselines.

\vspace*{-0.5\baselineskip}
\paragraph{Summary of differences}
There are several aspects that distinguish our work from relevant existing approaches. 
First, prior work has limited the task of value finding to that of identifying a single best value.  Conflicting data, however, can coexist~\citep{Vrandecic:2014:WFC}.  Casting value finding as a ranking problem provides a mechanism to deal with this plurality.
Second, most prior work relies on a single source, a table corpus, for finding missing values~\citep{Yakout:2012:IEA,Zhang:2013:ISM,Ahmadov:2015:THI}.  We also use a knowledge base, in addition, in case there are no relevant tables for the target entity and attribute on the Web.  The combination of a table corpus and a knowledge base has already been exploited in~\citep{Zhang:2018:OTG}. There, however, only the single best source is used.  Instead, we combine evidence from the two sources. 
Third, existing work either does not consider the data types of table columns~\citep{Zhang:2018:OTG,Ahmadov:2015:THI, Yakout:2012:IEA} or is limited to numerical values~\citep{Zhang:2013:ISM,Ahmadov:2015:THI}. We consider multiple value types, including non-numerical ones (entities and strings).

\vspace*{-0.5\baselineskip}
\section{Problem Statement and Overview}
\label{sec:ps}

We address the task of automatically finding the values of cells in relational tables.  
A table is said to be relational if it describes a set of entities in its core column (typically, the leftmost column) and the attributes of those entities in additional columns.  We shall assume that entities in the core column of each table have been identified and linked to a knowledge base.  
These annotations may be supplied manually (e.g., tables in Wikipedia) or can be obtained automatically~\citep{Efthymiou:2017:MWT,Ritze:2015:MHT}.
The additional (attribute) columns are identified by their heading labels. 

Formally, given an input table $T$, we seek to find the value of the cell that is identified by the row with entity $e$ (in the core column) and the column with heading label $h$.  
The output is a ranked list of values $v$, where the ranking of values is defined by a scoring function $\mathit{score}(v;e,h,T)$.

Our approach, shown in Fig.~\ref{fig:pipeline}, has two main components. First, we identify candidate values from two sources, a table corpus and a knowledge base.  Second, these values are ranked based on their likelihood of being correct, and the top-$k$ ranked values are presented to the user as auto-complete suggestions.  Note that it is a design decision for us to keep the user in the loop and let her make a judgment call on the appropriateness of a suggestion by considering the supporting evidence. 

The two main components of our  \textsc{CellAutoComplete} framework are described in the following two sections.


%

\section{Candidate Value Finding}
\label{sec:cf}

In this section, we address the problem of identifying candidate values for a given target cell in table $T$, identified by the target entity $e$ and target heading label $h$.  Candidate value finding is a crucial step as the recall of the end-to-end task critically depends on it. 
We gather candidate values from two sources: a table corpus (Sect.~\ref{sec:cf:tc}) and a knowledge base (Sect.~\ref{sec:cf:kb}).
Novel contributions in this part include the heading-to-heading and heading-to-predicate matchings, and the TMatch table matching approach.



\subsection{Table Corpus}
\label{sec:cf:tc}

Our goal is to locate tables from the corpus that contain the target entity and attribute (heading label) pair.  We assume a setting where entities in the core table columns  have been linked to a knowledge base (cf. Sect.~\ref{sec:expsetup:data}).  With that, it is easy to find the tables that contain the target entity (with high confidence).
The matching of heading labels, however, is not that straightforward: the same meaning may be expressed using different labels (e.g., ``established'' vs. ``founded''), while the same label can mean different things depending on the table's context (e.g., the column label ``played'' may refer, among others, to the number of games played, to the date of a game, or the name of the opponent).
Therefore, we need to perform a matching between heading labels (\S\ref{sec:cf:h2h}).
Naively considering all candidate tables that mention the entity and heading is not sufficient; additionally, we should also take into account their semantic similarity to the input table, referred to as the problem of \emph{table matching} (\S\ref{sec:cf:tablematching}).


\subsubsection{Heading-to-Heading Matching}
\label{sec:cf:h2h}

For a given heading label $h$, we wish to identify additional heading labels that have the same meaning (i.e., refer to the same entity attribute). 
This is closely related to the problem of schema matching~\citep{Dong:2005:MSA}.  
The main idea is that if two tables $T_a$ and $T_b$ contain the same value $v$ for a given entity $e$ in columns with headings $h_a$ and $h_b$, respectively, then $h_a$ and $h_b$ \emph{might} mean the same.  
Sharing a value does not always mean the equivalence of heading labels.  Nevertheless, the intuition is that the more often it happens, the more likely it is that $h_a$ and $h_b$ refer to the same entity attribute.  We capture this intuition in the following formula:
\begin{equation}
	P(h'|h) = \frac{n(h',h)}{\sum_{h''}n(h'',h)} ~, \label{eq:Phh}
\end{equation}
where $n(h',h)$ is the number of table pairs in the table corpus that contain the same value for a given entity in columns $h'$ and $h$, respectively.


\subsubsection{Table Matching}
\label{sec:cf:tablematching}

All tables in the table corpus that contain (i) the target entity $e$ and (ii) the target heading $h$ or any related heading label $h'$ ($n(h',h) > 0$), are considered as candidates.  As we explained above, not all these tables are actually good candidates.
Therefore, we estimate the semantic similarity between the input table $T$ and a candidate table $T'$, $\mathit{score}(T, T')$.  This table matching score later will be utilized as a confidence estimate in a subsequent value ranking step (in Sect.~\ref{sec:vr}).
We present two feature-based learning methods for table matching.  We start with a state-of-the-art approach, InfoGather. Then, we introduce TMatch, which extends InfoGather with a rich set of features from the literature.

\paragraph{InfoGather}

InfoGather~\citep{Yakout:2012:IEA} measures element-wise similarities across four table elements (table data, column values, page title, and heading labels), and combines them in a linear fashion:
\begin{equation*}
	\mathit{score}(T, T') = \sum_{x} w_x \times \mathit{sim}(T_x,{T'}_x) ~,
\end{equation*}
where $x$ refers to a given table element. Each table element $T_x$ is expressed as a term vector. Element-wise similarity $\mathit{sim}()$ is computed using the cosine similarity between the respective term vectors of the input and candidate tables.
%
%

\paragraph{TMatch}

\begin{table}[t]
\centering
\small
\caption{Overview of table ranking features used in TMatch.}
\vspace*{-0.75\baselineskip}
\label{tbl:features-tr}
\begin{tabular}{p{0.1cm}p{6.5cm}l}
	\toprule
	\multicolumn{2}{l}{\textbf{Group / Feature}} & \textbf{Source} \\ 
	\midrule
	\multicolumn{3}{l}{\emph{Table features}} \\
	& Number of rows in the table & \cite{Cafarella:2008:WEP,Bhagavatula:2013:MEM} \\
	& Number of columns in the table & \cite{Cafarella:2008:WEP,Bhagavatula:2013:MEM}  \\
	& Number of empty table cells & \cite{Cafarella:2008:WEP,Bhagavatula:2013:MEM} \\
	& Table caption IDF & \cite{Qin:2010:LBC} \\	
	& Table page title IDF & \cite{Qin:2010:LBC} \\	
	& Number of in-links to the page embedding the table & \cite{Bhagavatula:2013:MEM} \\
	& Number of out-links from the page embedding the table & \cite{Bhagavatula:2013:MEM} \\
	& Number of page views & \cite{Bhagavatula:2013:MEM} \\
	& Inverse of number of tables on the page & \cite{Bhagavatula:2013:MEM} \\
	& Ratio of table size to page size & \cite{Bhagavatula:2013:MEM} \\
	\midrule
	\multicolumn{3}{l}{\emph{Matching features}} \\ 	
	& InfoGather page title IDF similarity score & \cite{Yakout:2012:IEA} \\
	& InfoGather heading-to-heading similarity & \cite{Yakout:2012:IEA} \\
	& InfoGather column-to-column similarity & \cite{Yakout:2012:IEA} \\
	& InfoGather table-to-table similarity & \cite{Yakout:2012:IEA} \\	
	& MSJE heading matching score & \cite{Lehmberg:2015:MSJ} \\
	& \citet{Nguyen:2015:RSS} heading similarity & \cite{Nguyen:2015:RSS} \\
	& \citet{Nguyen:2015:RSS} table data similarity & \cite{Nguyen:2015:RSS} \\
	& Schema complement schema benefit score & \cite{DasSarma:2012:FRT} \\
	& Schema complement entity overlap score &  \cite{DasSarma:2012:FRT} \\
	& Entity complement entity relatedness score & \cite{DasSarma:2012:FRT}  \\  
	\bottomrule
\end{tabular}
\vspace*{-1\baselineskip}
\end{table}


We extend the four element-wise matching scores of InfoGather with a number of additional matching, which are summarized in Table~\ref{tbl:features-tr}.  We use Random Forests regressor as our machine-learned model.
We distinguish between two main groups of features. The first group of features (top block in Table~\ref{tbl:features-tr}) aim to characterize an individual table and are associated with its quality and importance.
These features are computed for both the input and candidate tables.
The second group of features (bottom block in Table~\ref{tbl:features-tr}) measures the degree of matching between the input and candidate tables.  
In the interest of space, we present a high-level description of these features and refer to the original publications for details.

\begin{itemize}
	\item The Mannheim Search Join Engine (MSJE)~\citep{Lehmberg:2015:MSJ} measures the similarity between the headings of two tables by creating an edit distance graph between the input and candidate tables' heading terms. 
		Then, the \emph{maximum weighted bipartite matching score} is computed on this graph's adjacency matrix.
	\item \citet{Nguyen:2015:RSS} consider the table headings and table data for matching. Specifically, heading similarity is computed by solving the \emph{maximum weighted bipartite sub-graph problem}~\cite{Anan:2007:OSM}.  Data similarity is measured by representing each table column as a binary term vector, and then taking the cosine similarity between the most similar column pairs.
	\item \citet{DasSarma:2012:FRT} compute the matching score by aggregating the  benefits of adding additional headings columns and entities from the candidate table to the input table.
\end{itemize}

\subsection{Knowledge Base}
\label{sec:cf:kb}

Next, we discuss how to utilize a knowledge base for gathering candidate values.  By definition, the columns in relational tables correspond to entity attributes.  The main challenge to be addressed here is how to map the column heading label to the appropriate KB predicate (\S\ref{sec:cf:h2p}).  After that, candidate values can easily be fetched from the KB (\S\ref{sec:cf:ve}).

\subsubsection{Heading-to-Predicate Matching}
\label{sec:cf:h2p}

\begin{figure}[t]
\vspace*{-1\baselineskip}
   \centering
   \includegraphics[width=0.5\textwidth]{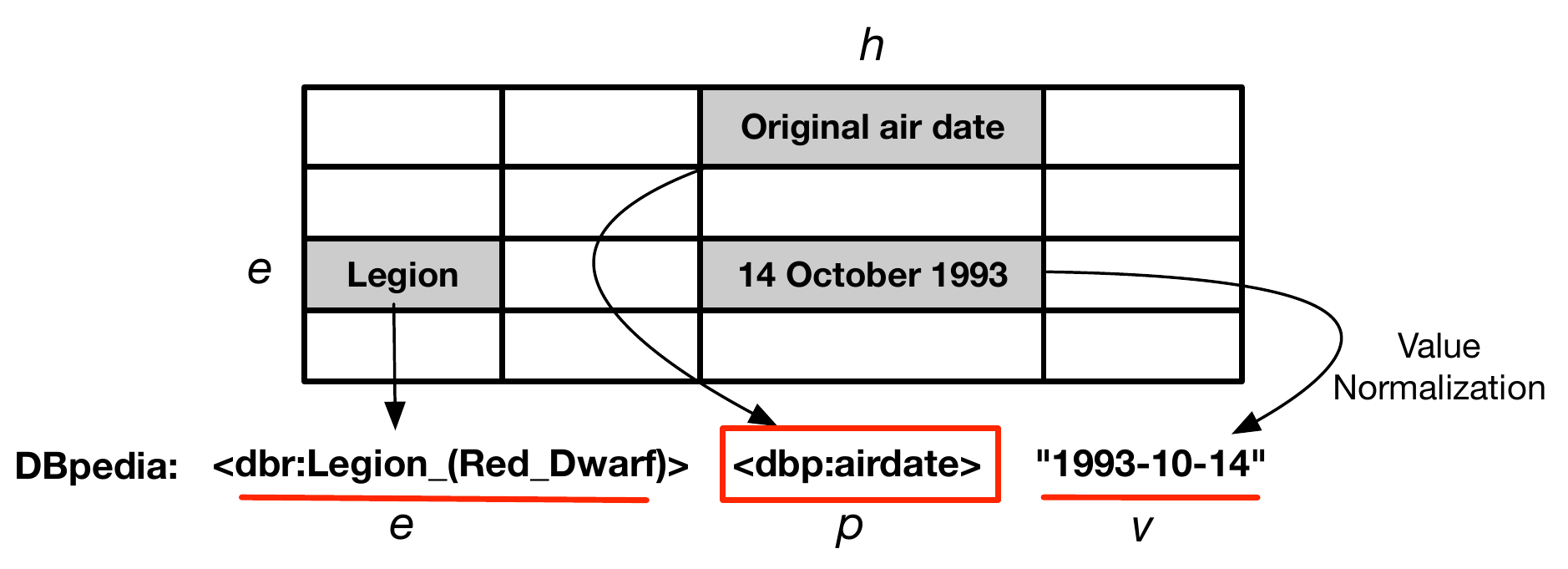} 
   \vspace*{-1.5\baselineskip}
   \caption{Illustration of table heading label to KB predicate matching.  Notice that value normalization is also involved.}
   \label{fig:hp}
\end{figure}

Both table heading labels and knowledge base predicates represent entity attributes, but these are often expressed differently, making string matching insufficient.    
Similarly to how it was done in heading-to-heading matching, we capitalize on the observation that if entity $e$ has value $v$ for predicate $p$ in the KB, and the same entity has value $v$ in the heading column $h$ of many tables, then $p$ and $h$ are likely to mean the same (more precisely, $h$ is a string label that corresponds to the semantic relation $p$).  This idea is illustrated in Fig.~\ref{fig:hp}.  
The similarity between heading label ($h$) and predicate ($p$) is computed according to the conditional probability $P(p|h)$:
\begin{equation}
	P(p|h) = \frac{n(h,p)}{\sum_{p'}n(h, p')} ~, \label{eq:Pph}
\end{equation}
where $n(h,p)$ denotes the times of $h$ and $p$ indicate the same value in the corpus. 
Table~\ref{tbl:tmp} lists some examples.
\begin{table}[t]
	\centering
	\footnotesize
	\caption{Examples of heading label and predicate matches.}
   \vspace*{-1\baselineskip}
	\begin{tabular}{lll}
		\toprule
		Heading label ($h$) & Predicate ($p$) & $n(h,p)$ \\
		\midrule
		Director  & dbp:director & 38587 \\
		& dbp:writer & 2348 \\
		Location & dbp:city & 10772 \\
		& dbp:location & 9170 \\
		Country & dbp:birthPlace & 9077 \\
		 & dbp:country & 3546 \\		
		 
		
		\bottomrule
	\end{tabular}
	\label{tbl:tmp}
\end{table}

\subsubsection{Value Extraction}
\label{sec:cf:ve}

Given a table heading, all matching predicates (i.e., where $n(h,p)>0$) are considered.
Then, for each of these predicates $p$, the object values associated with the subject $e$ and predicate $p$ in the knowledge base are considered as candidate values (i.e., all the subjects of SPO triples matching the pattern $\langle e, p, ? \rangle$).

\section{Value Ranking}
\label{sec:vr}

In this section we describe methods for ranking the candidate cell values that were identified in the previous step.  
For each source, we have a set of candidate values $V$, and for each of the candidate values $v \in V$ a set of supporting evidence sources $S_v$.  In the case of a knowledge base, the triples where the predicate matches the target heading $h$ or related headings $h'$ are the evidence sources.  In the case of a table corpus, $S_v$ contains all candidate tables $T'$ where $v$ is a cell value corresponding the target entity $e$ and to heading label $h'$.
Our task is to score each candidate value $v$ based on the available evidence.

There are two main challenges we need to deal with. One is how to handle \empt values, i.e., quantify our confidence in that the given cell should be left empty.
Another is how to combine evidence across multiple sources, specifically, a knowledge base and a table corpus.
We start by considering each source individually in Sects.~\ref{sec:vr-kb} and~\ref{sec:vr-tc}, and then combine the two in a feature-based learning approach in Sect.~\ref{sec:vr-comb}.
Novel contributions in this part include the source-specific value finding methods as well as the three groups of features, each designed with a specific intuition in mind: (i) quantifying the support each source has for a given value, (ii) dealing with \empt values, and (iii) effectively prioritizing values from semantically more related tables in a table corpus.

\subsection{Knowledge Base}
\label{sec:vr-kb}

We deal with empty values by adding a designated special value \empt to the set of candidates.  The scoring of values is then based on the following formula:
\begin{align}
   \mathit{score}(v;e,h,T) =
     \left \{ \begin{array}{ll}
          \arg\max_{p} \mathit{score}(p,h), ~~~~ & v \neq \text{\empt} \\
          \gamma, & v = \text{\empt} ~,
          \end{array}
     \right .
     \label{eq:score-kb}
\end{align}
where $\gamma$ is a free parameter that we learn empirically.  For non-empty values, $score(p,h)$ can be estimated in two alternative ways:

\begin{itemize}
	\item We take the edit distance between the column heading and the (label of the) predicate\footnote{We are not using the predicate itself, but the corresponding label from the KB that is meant for human consumption. E.g., for \texttt{<dbp:timeZone>} the label is ``time zone.''} (referred to as soft matching in~\citep{Zhang:2018:OTG}).  
		\begin{equation}
			\mathit{score}_{ED}(p,h) = 1 - \frac{\mathit{dist}(p,h)}{\max(|p|,|h|)} ~,	 \label{eq:ed}
		\end{equation}
		where $\mathit{dist}()$ represents the minimum number of single-cha\-rac\-ter edit operations (i.e., insertion, deletion, or substitution) needed to transform one string into another. 
		
	\item We use the conditional probability $P(p|h)$ (cf. Eq.~\eqref{eq:Pph}).
\end{itemize}

\subsection{Table Corpus}
\label{sec:vr-tc}


A given candidate value may have multiple supporting tables in the corpus.  We formulate two evidence combination strategies.
One method (\S\ref{sec:vr:topr}) is to consider a single table that best matches semantically the input table, similarly to~\cite{Ahmadov:2015:THI}.
Another method (\S\ref{sec:vr:all}) is to consider multiple tables but weigh them according on their semantic similarity to the input table, an idea in line with~\citep{Yakout:2012:IEA,Zhang:2013:ISM}.
Both methods are based on the notion of \emph{table matching}, which we described in \S\ref{sec:cf:tablematching}.
One important addition, compared to prior approaches, is that we also consider which heading $h'$ of the candidate table $T'$ matches best the target heading $h$ (\S\ref{sec:vr:hls}).  As we will show in our experiments, this has a positive effect.



\subsubsection{Top-ranked Table}
\label{sec:vr:topr}

This method takes the best matching candidate table $T'$, i.e., the one that is most similar to the input table $T$, and combines that table's matching score with the best matching heading label within that table.
Formally:
\begin{equation}
	score(v,e,h,T) = \arg\max_{T'} score(T',T) \times \big ( \max_{h' \in T'} \mathit{sim}(h',h) \big ) ~. \label{eq:tscore_top}
\end{equation} 
%

\subsubsection{All Tables}
\label{sec:vr:all}


Alternatively, one might consider all matching tables, as opposed to a single best one, and aggregate information from these.
Formally:
\begin{equation}
	\mathit{score}(v,e,h,T) = \sum_{T'} \Big ( \mathit{score}(T',T) \times  \max_{h' \in T'} \mathit{sim}(h',h) \Big ) ~, \label{eq:tscore_all}
\end{equation}
where $\mathit{sim}(h',h)$ is the similarity between two heading column labels, which we detail below.

\subsubsection{Heading Label Similarity}
\label{sec:vr:hls}

 We consider four methods for computing the similarity $\mathit{sim}(h',h)$ between two heading column labels: 
 \begin{itemize}
	\item \emph{Uniform}: we set similarity to a fixed value (e.g., 1). This way the similarity of headings is not considered at all in Eqs.~\eqref{eq:tscore_top} and~\eqref{eq:tscore_all}. (The uniform estimator will merely serve as a baseline, to evaluate the benefits of incorporating heading similarity.)
 	\item \emph{Edit distance}: We use the edit distance between $h$ and $h'$ (as in Eq.~\eqref{eq:ed}, but replacing $p$ with $h'$).
 	\item \emph{Mapping probability}: we use the conditional probability $P(h'|h)$ as defined in Eq.~\eqref{eq:Phh}.
 	\item \emph{Label2Vec}: We employ the skip-gram model of Word2vec~\citep{Mikolov:2013:DRW} to train heading label embeddings on the table corpus.  Then, $\mathit{sim}(h',h)$ is taken to be the cosine similarity between the embedding vectors of $h'$ and $h$, respectively.
 \end{itemize}
 
\begin{table*}[t]
\centering
\small
\caption{Features for value ranking. $e$ and $h$ denote the entity and heading column, respectively in the input table $T$, while $h'$ is the best matching column (based on mapping probability) in the candidate table $T'$.}
\vspace*{-0.75\baselineskip}
\begin{tabular}{p{0.05cm}p{4.6cm}p{9.5cm}ll}
	\toprule
	\multicolumn{2}{l}{\textbf{Group / Feature}} & \textbf{Description} & \textbf{Source} & \textbf{Value} \\ 
	\midrule
	\multicolumn{4}{l}{\emph{Feature group I}} \\ 	
	& IS\_TC & Whether the value comes from the table corpus ($v \in V_{TC}$) & TC & $\{0,1\}$ \\
	& IS\_KB & Whether the value comes from the knowledge base ($v \in V_{KB}$) & KB & $\{0,1\}$ \\
	& EDITDIST\_PH & Predicate-to-heading edit distance ($\mathit{score}_{ED}(p,h)$) & KB, TC & $[0,1]$\\
	& MAPPINGPROB\_PH & Predicate-to-heading mapping probability ($P(p|h)$) & KB, TC & $[0,1]$\\	
	& EDITDIST\_HH & Heading-to-heading edit distance ($\mathit{score}_{ED}(h',h)$) & TC & $[0,1]$\\
	& MAPPINGPROB\_HH & Heading-to-heading mapping probability ($P(h'|h)$) & TC & $[0,1]$\\
	\midrule
	\multicolumn{4}{l}{\emph{Feature group II}} \\
	& NUM\_E & Number of times entity $e$ appears in the table corpus & TC & $[0,\infty)$ \\ 
	& NUM\_H & Number of times heading $h$ appears in the table corpus & TC & $[0,\infty)$ \\ 
	& NUM\_EH & Number of times entity $e$ and heading $h$ co-occur in the table corpus & TC & $[0,\infty)$ \\ 
	& EMPTY\_RATE & Fraction of empty cells in column $h$ in the table corpus & TC & $[0,1]$ \\
	& MATCH\_PH\_NUM & Number of predicate-to-heading matches ($|\{p:P(p|h) > 0\}|$) & KB, TC & $[0,\infty)$ \\
	& MATCH\_HH\_NUM & Number of heading-to-heading matches ($|\{h': P(h'|h) > 0\}|$) & TC & $[0,\infty)$ \\
	& MATCH\_PH\_\{MAX,AVG,SUM\} & Aggregated number of predicate-to-heading matches ($\mathit{aggr}_{p}
	\big[\mathit{count}(p,h)\big]$) & KB, TC & $[0,\infty)$ \\
	& MATCH\_HH\_\{MAX,AVG,SUM\} & Aggregated number of heading-to-heading matches ($\mathit{aggr}_{h'}
	\big[\mathit{count}(h',h)\big]$) & TC & $[0,\infty)$ \\
	\midrule
	\multicolumn{4}{l}{\emph{Feature group III}} \\
	& TMATCH\_NUM & Number of matching tables, using TMatch scorer ($|\{T': \mathit{score}_{HCF}(T',T) > 0\}|$) & TC & $[0,\infty)$ \\
	& TMATCH\_\{MAX,AVG,SUM\} & Aggregated table matching scores, using TMatch scorer ($\mathit{aggr}_{T'}
	\big[\mathit{score}_{HCF}(T',T)\big]$) & TC & $[0,\infty)$ \\
	& SCORE\_IG\_ED\_\{MAX,AVG,SUM\} & Aggregated value score using InfoGather table matching with edit distance & TC & $[0,\infty)$ \\
	& SCORE\_IG\_MP\_\{MAX,AVG,SUM\} & Aggregated value score using InfoGather table matching with mapping probability & TC & $[0,\infty)$ \\
	& SCORE\_IG\_L2V\_\{MAX,AVG,SUM\} & Aggregated value score using InfoGather table matching with Label2vec & TC & $[0,\infty)$ \\
	& SCORE\_TMATCH\_ED\_\{MAX,AVG,SUM\} & Aggregated value score using TMatch table matching with edit distance & TC & $[0,\infty)$ \\
	& SCORE\_TMATCH\_MP\_\{MAX,AVG,SUM\} & Aggregated value score using TMatch table matching with mapping probability & TC & $[0,\infty)$ \\
	& SCORE\_TMATCH\_L2V\_\{MAX,AVG,SUM\} & Aggregated value score using TMatch table matching with Label2vec & TC & $[0,\infty)$ \\
	\bottomrule
\end{tabular}
\vspace*{-0.5\baselineskip}
\label{tbl:features-vr}
\end{table*}

\subsection{Combination of Evidence}
\label{sec:vr-comb}

We combine evidence from multiple sources using a feature-based approach.  Table~\ref{tbl:features-vr} summarizes our features.
Additionally, we use the same table quality/importance features as for table matching (cf. top block in Table~\ref{tbl:features-tr}).

Feature group I captures how much support there is for the given value in each source. 
Two binary features (IS\_TC and IS\_KB) are meant to indicate whether the value can be found in a given source (table corpus and knowledge base). 
The next four features are used for capturing heading level similarity, based on edit distance (EDITDIST\_PH and EDITDIST\_HH) and mapping probability (MAPPINGPROB\_PH and MAPPINGPROB\_HH). 

Feature group II aims at empty value prediction. The intuition is that if the entity-heading combination appears a lot in the table corpus or in the knowledge base, then we have a better chance of finding a value. 
Some features (NUM\_E, NUM\_H, and NUM\_EH) are general statistics on the number of entity, heading or entity-heading occurrences in the tables corpus.
EMPTY\_RATE measures the fraction of cells that are empty in a given column.
MATCH\_EH\_NUM and MATCH\_HH\_NUM are the number of predicate-heading and heading-heading matches. 
The last 6 features are the aggregated counts of predicate-heading and heading-heading matches in the knowledge base and in the table corpus.\footnote{For predicate-to-heading and heading-to-heading matching, empty values are not considered, i.e., two empty cell values are not considered as being the same.}  


Feature group III aims for capturing the semantic relatedness between the input table and candidate table where the value is taken from.
The matches between the input table and candidate tables are captured in the number of matching tables (TMATCH\_NUM) as well as aggregates over the table matching scores (TMATCH\_*).
Additionally, we consider the value scoring mechanism devised specifically for TC (cf. Sect.~\ref{sec:vr-tc}), which involves a table matching method (InfoGather (IG) or TMatch), heading similarity (edit distance (ED), mapping probability (MP), or Label2vec (L2V)), and an aggregator (MAX, AVG, or SUM).  All possible combinations yield a total of 18 features.  For example, SCORE\_IG\_ED\_SUM corresponds to Eq.~\eqref{eq:tscore_all} using InfoGather for table matching and edit distance heading similarity, and SCORE\_TMATCH\_L2V\_MAX corresponds to Eq.~\eqref{eq:tscore_top} using TMatch table matching, and Label2Vec heading similarity.

\vspace*{-0.5\baselineskip}
\section{Experimental Setup}
\label{sec:expsetup}

Auto-completion for data cells is a novel problem, and as such, no public test collection exists.  In this section, we introduce the data sources used in our experiments and describe the construction of our test collection, which is another main contribution of this study. It is based on 1000 table cells and contains labels for 35k cell-value pairs, obtained via crowdsourcing.
We also present the techniques we employed for column data type detection and value normalization, which are essential to ensure data quality.

\vspace*{-0.5\baselineskip}
\subsection{Data Sources}
\label{sec:expsetup:data}

We use two main data sources:

\begin{itemize}
	\item \textbf{Table Corpus (TC)} The WikiTables corpus~\citep{Bhagavatula:2015:TEL} is extracted from Wikipedia and contains 1.6M high-quality tables.  Following~\citep{Zhang:2018:AHT}, we select the core column by taking the one among the left-most 2 columns with the highest entity rate.  Based on a sample of 100 tables, this method has over 98\% accuracy.	There are 755k relational tables in the corpus that have a core column where 80\% of the cell values are entities.  Since tables are from Wikipedia, the mentioned entities have been explicitly marked up.
	\item \textbf{Knowledge Base (KB)} DBpedia is a general-purpose knowledge base that is a central hub in the Linked Open Data cloud.  Specifically, we use the 2015-10 version and restrict ourselves to entities that have at least a short description (abstract), amounting to a total of 4.6M entities.
\end{itemize}


\vspace*{-0.5\baselineskip}
\subsection{Column Data Type Detection}


Our objective is to classify a given table column according to some taxonomy of data types.  It is assumed that all cells within a column share the same data type.
To determine the data type of a given column, we classify each (non-empty) cell within that column and then take a majority vote.  In the rare case of a tie, the column will be assigned multiple types.
In the following, we introduce the value data type taxonomy used and our method for classifying the value data types of individual table cells. 

\begin{figure}[t]
   \centering
   \vspace*{-1\baselineskip}   
   \includegraphics[width=0.48\textwidth]{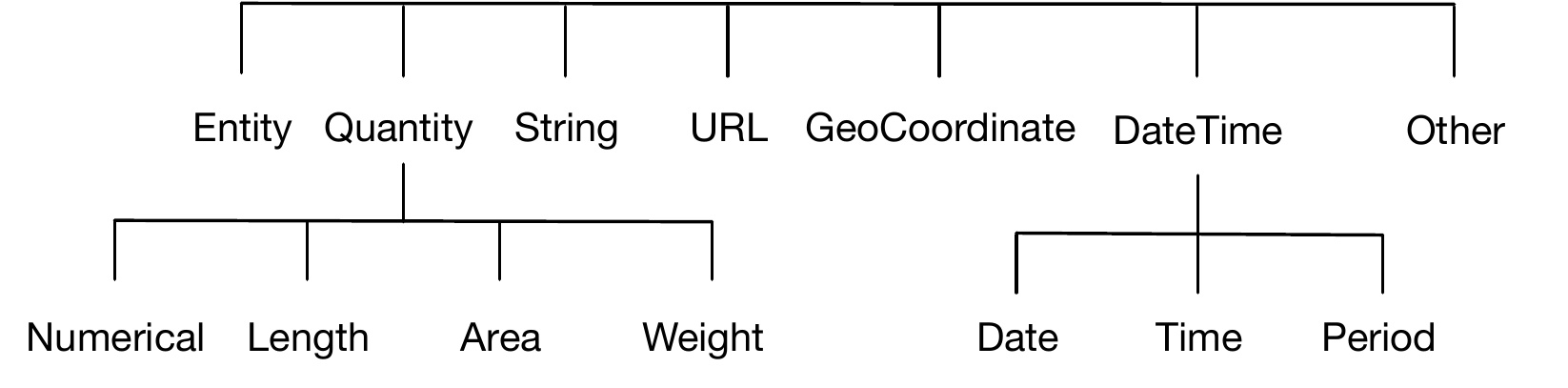} 
   \vspace*{-1.5\baselineskip}
   \caption{Value data type taxonomy used in this paper.}
	\vspace*{-0.5\baselineskip}   
   \label{fig:type_taxonomy}
\end{figure}

\subsubsection{Value Data Type Taxonomy}
Different value data type taxonomies have been proposed in the literature, see, e.g.,~\citep{Ritze:2015:MHT,Yin:2011:FFL}. 
We build on and extend the taxonomy by \citet{Yin:2011:FFL}, who consider seven value data types in the context of fact finding from web tables: string, date/time, numerical, duration, length, area, and weight. 
Knowledge bases also have their own data type taxonomies, e.g., 
DBpedia has 25 data types.\footnote{http://mappings.dbpedia.org/index.php/DBpedia\_Datatypes} 
Informed by these, we introduce a two-layer value data type taxonomy, which is shown in Figure~\ref{fig:type_taxonomy}.
We manually map the data types of the knowledge base (here: DBpedia) to our value data type taxonomy.

\subsubsection{Cell Value Type Classification}
\label{sec:cf:vtclass}

Following standard practice~\citep{Ritze:2015:MHT}, we design a rule-based method for classifying cell values into our value data type taxonomy.
For example, DateTime values are identified based on the cell values matching given patterns and on certain terms appearing in the column heading label (such as ``year,'' ``birth,'' ``date,'' ``founded,'' ``created,'' or ``built''). The complete set of rules is released in the online appendix.

\subsubsection{Evaluation}
We obtain the distribution of column data types based on a sample of 100k relational tables from the table corpus. 
The results, ordered by frequency, are as follows: 
(1) Quantity: 269,260 cells (43.2\%), 
(2) Entity: 200,637 cells (32.2\%),
(3) String: 76,259 cells (12.2\%),
(4) Other: 51,999 cells (8.3\%),
(5) DateTime: 24,413 cells 3.9\%,
(6) GeoCoordinate: 161 cells (0.0\%).
There were not any cells of type URL, as in our sample all links refer to entities in Wikipedia.
Given that number of columns with type GeoCoordinate is negligible, we exclude this type in our experiments.  We further note that columns with type Other contain mostly empty values.

To verify whether the performance of our column type detection method is sufficient, we manually evaluate it on a sample of 100 tables.  Specifically, tables are selected such that each has at least 6 rows and 4 columns, and has a core column where over 80\% of the cell values are entities.  
Our sample contains a total of 473 table columns.  The accuracy of column type detection is found to be 94.92\%.

%
%

\vspace*{-0.5\baselineskip}
\subsection{Value Normalization}
\label{sec:sub:vp}

The previous step informs us about the data type of the value that we are looking for.  
Values, however, may be expressed in a variety of ways in different sources.  For example, dates are written differently by individuals in different parts of the world.
We normalize cell values according to the data types of the corresponding column.  


To ensure high data quality, we employ a rule-based approach.  On close inspection of the data, we develop over 100 rules for normalizing cell values based on their data types.  We illustrate these transformations with some examples. 
All dates are converted to ``YYYY-MM-DD'' format and all times are transformed to ``HH:MM:SS'' format. 
Date periods with only years are normalized to ``year--year'' and those with dates are separated into two dates. E.g., ``1998--99'' is normalized to ``[1998,1999],'' while ``5 October 1987 to 30 December 1987'' is converted to ``[1987-10-05, 1987-12-30].''
For quantities, the numeric values and the units are kept separately, e.g, ``100 m'' is stored as (100, ``m'') and ``-54 kilograms'' is stored as (-54, ``kilograms'').
No unit conversion is performed.
In the case of composite values, we only keep the first value, e.g., ``71 kg/m$^2$ (14.5 lb/ft$^2$)'' is stored as (71, ``kg/m$^2$'').

\vspace*{-0.5\baselineskip}
\subsection{Test Collection}

\begin{figure}[t]
\vspace*{-1\baselineskip}
   \centering
   \includegraphics[width=0.45\textwidth]{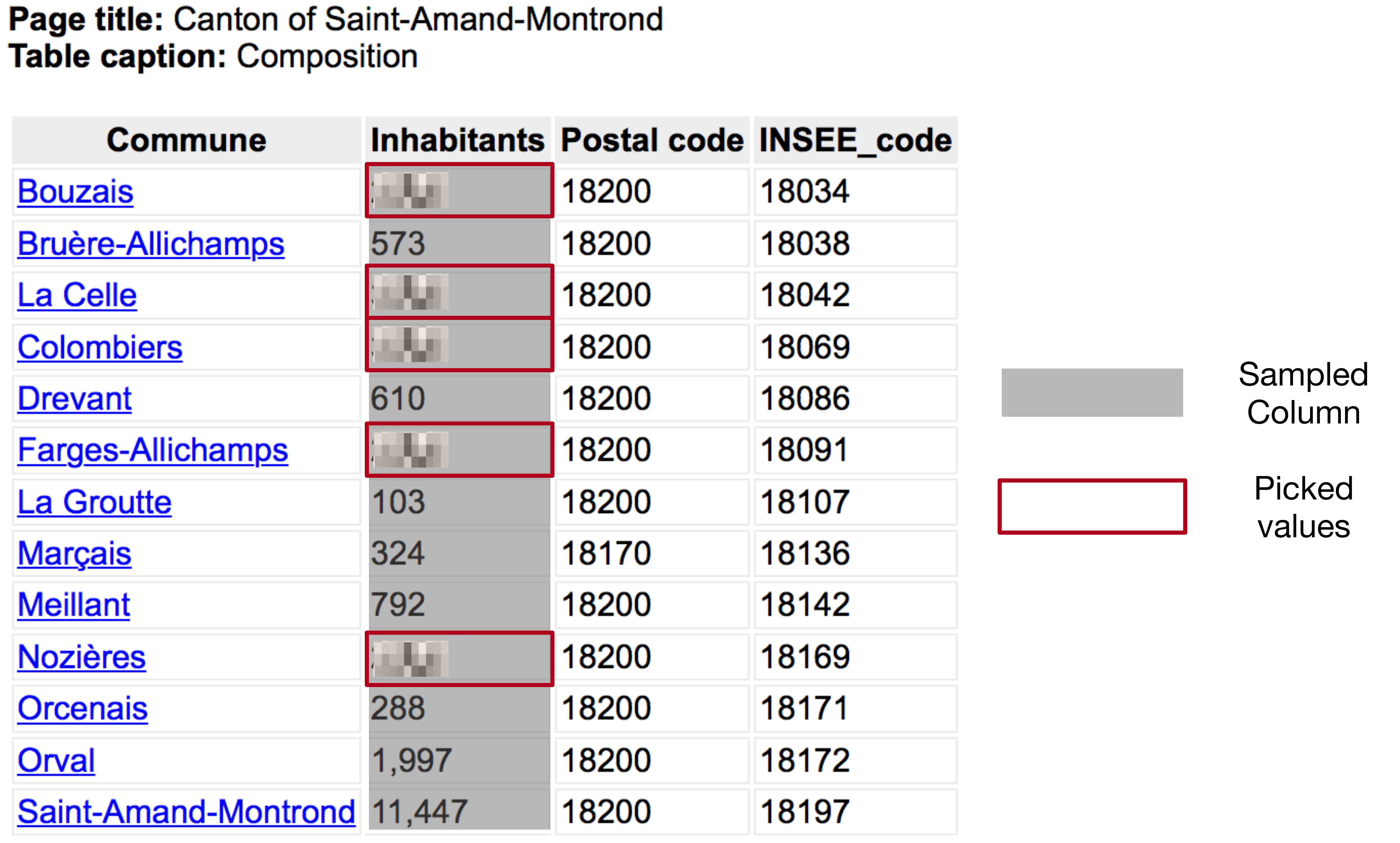} 
   \vspace*{-0.5\baselineskip}
   \caption{Illustration of sampling values. First, the column of ``Inhabitants'' is the sampled as a numerical column. Then, five values in this column are picked randomly as test cases.}
   \label{fig:cs}
   \vspace*{-0.5\baselineskip}
\end{figure}

We create a test collection for value finding based on a sample of existing tables from the table corpus. (These test tables are excluded from our index and when computing statistics.)
Specifically, we perform stratified sampling according to the four main column data types: Entity, Quantity, String, and DateTime.
For each data type, we first randomly select 50 columns, each from a different table, where there is at least 80\% agreement on the column data type according to the majority vote method (cf. Sect.~\ref{sec:cf:vtclass}).  We further require that the table has at least 5 rows and 3 columns, and the respective heading label has a certain minimum length (4 characters).  From each sampled table column, 5 specific cells are picked randomly.  This way, our test collection consists of $4 \times 50 \times 5=1000$ cells for which we are trying to find values.  These input tables are then excluded from the collection.
See Fig.~\ref{fig:cs} for an illustration.

%
%
%
%

%
\begin{table}[t]
	\centering
	\caption{Test collection statistics (based on 1000 cells).}
	\label{table:gt}
	\vspace*{-0.5\baselineskip}
	\begin{tabular}{l|cc|c}
		\toprule
		 & KB & TC & KB+TC \\
		\midrule
		Avg. \#values & 1.31 & 2.51 & 2.65 \\
		\empt rate & 0.48 & 0.31 & 0.18 \\
		\bottomrule
	\end{tabular}
	\vspace*{-0.5\baselineskip}
\end{table}
%

%
%
\begin{table*}[t]
	\centering
	\caption{Value finding performance. Significance for line $i$ ($i>1$) is tested against the best method in lines $1..i-1$. }
	\label{table:vf-kb+tc}
	\vspace*{-0.75\baselineskip}
	\begin{tabular}{llccllll}
		\toprule
		Source & Method & \multicolumn{2}{c}{Sources used} & \multicolumn{2}{c}{\empt excluded}  & \multicolumn{2}{c}{\empt included} \\
		 & & KB & TC & NDCG@5 & NDCG@10 & NDCG@5 & NDCG@10 \\
		\midrule
		\emph{Single-source}
		& KBLookupED & \checkmark & & 0.2635 & 0.2652 & 0.2780 & 0.2806 \\
		& InfoGather, top, UNI  & & \checkmark & 0.4563$^\ddag$  & 0.4710$^\ddag$ & 0.4158$^\ddag$ & 0.4302$^\ddag$  \\ 
		& InfoGather, top, L2V  & & \checkmark & 0.4868  & 0.4978 & 0.4413 & 0.4537  \\
		& TMatch, top, UNI  & & \checkmark & 0.4744$^\ddag$  & 0.4873$^\ddag$ & 0.4297$^\ddag$ & 0.4417$^\ddag$   \\ 
		& TMatch, top, L2V  & & \checkmark & 0.5046  & 0.5139 & 0.4531 & 0.4624  \\ 
		\midrule
		\emph{Multi-source} 
		& OTG~\citep{Zhang:2018:OTG} & \checkmark & \checkmark & 0.5856 & 0.6062 & 0.5185 & 0.5367 \\
		& \textsc{CellAutoComplete} (feat. I)  & \checkmark & \checkmark & 0.6641$^\ddag$   & 0.6826$^\ddag$  & 0.5766$^\ddag$  & 0.5954$^\ddag$    \\
		& \textsc{CellAutoComplete} (feat. I+II)  & \checkmark & \checkmark & 0.6844$^\ddag$   & 0.7034$^\ddag$  & 0.5905$^\ddag$  & 0.6100$^\ddag$    \\
		& \textsc{CellAutoComplete} (feat. I+II+III)  & \checkmark & \checkmark & \textbf{0.7570}$^\ddag$ & \textbf{0.7641}$^\ddag$ & \textbf{0.6716}$^\ddag$ & \textbf{0.6785}$^\ddag$ \\
		\bottomrule
	\end{tabular}
	\vspace*{-0.5\baselineskip}
\end{table*}

Relevance assessments were collected via crowdsourcing using the Figure Eight platform.\footnote{\url{https://www.figure-eight.com/}}
For each cell, human assessors were presented with the page title (embedding the table), table caption, the core column entity, the heading column label, and a source document.  The source document is either the DBpedia page of the core column entity or an existing table from the table corpus.  Users were then asked to check if the missing cell value can be found within the source document, and, if yes, to provide the corresponding value (otherwise enter a designated special \empt value).
Each instance was annotated by 7 assessors. 
The inter-annotator agreement in terms of Fleiss' kappa statistic was about 0.7 when using the knowledge base and 0.8 when using the table corpus as source.  The former is considered as substantial, the latter is considered as almost perfect agreement~\citep{Landis:1977:MOA}.
The total expense of the crowdsourcing experiments was \$770. 

We then combine the correct values from these two sources as our ground truth. (We only use the KB and TC specific subsets in our analysis of specific sources in Sect.~\ref{sec:eval:sources}.)
Table~\ref{table:gt} shows statistics of our test collection.
We find that, when using both sources, cells on average have over two possible correct values.  
It is further worth noting that the rate of empty cells is much lower when combining the two sources, attesting to their complementary nature.

\vspace*{-0.5\baselineskip}
\subsection{Table Matching}
To train our table matching models (InfoGather and TMatch in Sect.~\ref{sec:cf:tablematching}), we construct a training dataset.  We group tables by topics and sample 50 tables with diverse topics (such as military, paleontology, sports, geography, etc.) from the corpus as input tables.  Each table should have at least five rows and three columns. For each table, we utilize the query-based search methods in~\cite{Ahmadov:2015:THI} to obtain a set of candidate tables.  We ask 3 annotators to judge if the candidate table is highly relevant, relevant, or not relevant.  

\vspace*{-0.5\baselineskip}
\subsection{Evaluation Measures}

We evaluate performance in terms of Normalized Discounted Cumulative Gain (NDCG) at cut-off points 5 and 10.  
To test significance, we use a two-tailed paired t-test and write $\dag$/$\ddag$ to denote significance at the 0.05 and 0.01 levels, respectively.

\section{Experimental Evaluation}
\label{sec:eval}

This section presents evaluation results for the value finding task (Sect.~\ref{sec:eval:res}) followed by further analysis of value sources (Sect.~\ref{sec:eval:sources}), features (Sect.~\ref{sec:eval:anal}), and specific examples (Sect.~\ref{sec:eval:examples}).

\subsection{Evaluating Auto-Completion}
\label{sec:eval:res}

We begin with the evaluation of the end-to-end cell value auto-completion task. Table~\ref{table:vf-kb+tc} reports the results.
At the top block of Table~\ref{table:vf-kb+tc}, we display the methods that use an individual source, either knowledge base (KB, line 1) or table corpus (TC, lines 2--5).  These methods meant to serve as single-source baselines; they are further detailed in Sect.~\ref{sec:eval:sources}.  
The bottom block of Table~\ref{table:vf-kb+tc} shows methods that utilize both sources.  There is only one existing work in the literature that we found directly applicable: the On-the-Fly Table Generation (OTG) approach by \citet{Zhang:2018:OTG}. This method combines a knowledge base and a table corpus in a simple way, by always giving preference to the former source over the latter. 

Looking at the results in Table~\ref{table:vf-kb+tc}, it is clear that the table corpus is a more effective source for value finding than the knowledge base.  
At the same time, they are complementary and combining the two yields substantial improvements. This is already witnessed for OTG~\citep{Zhang:2018:OTG}, but to a much larger extent with our \textsc{CellAutoComplete} methods.
Our best methods, using the complete feature set (cf. Table~\ref{tbl:features-vr}) outperforms OTG substantially, i.e., by over 26\% on all evaluation metric and experimental conditions (lines 6 vs. 9). 
These improvements can be attributed to two main factors.
First, instead of naively giving preference to the knowledge base over the table corpus, as in OTG, \textsc{CellAutoComplete} (feat. I) decides for each cell individually which source should be preferred, by considering the predicate-to-heading and heading-to-heading matching probabilities, among other signals.  This makes a large difference, as can be observed in the scores (lines 6 vs. 7).
Second, taking into account the semantic similarity of tables, when using a table corpus as source, makes a large difference. This is what feature group III contributes. We find that it brings in an over 10\% relative improvement, see  \textsc{CellAutoComplete} (feat. I+II) vs. (feat. I+II+III), i.e., the bottom two lines in Table~\ref{tbl:features-vr}.
As for the second group of features, which aims at improving empty value prediction, we find that is has a small, but positive and significant impact (feat. I vs. I+II). 





\subsection{Analysis of Sources}
\label{sec:eval:sources}

Next, we analyze cell auto-completion performance using only a single source: a knowledge base (Table~\ref{table:vf-kb}) and a table corpus (Table~\ref{table:vf-tc}).
As before, we distinguish between two settings, with \empt values excluded and included.  We note that the ground truth is restricted to the specific source, therefore, it is different in the two cases (and also different from Table~\ref{table:vf-kb+tc}, which uses the union of the two).

\subsubsection{Using a Knowledge Base}

We compare two different KB-based value lookup methods, edit distance (ED) and matching probability (MP), in Table~\ref{table:vf-kb}.  The two methods yield virtually identical performance when \empt values are excluded.  When \empt values are considered, ED performs significantly better than MP.
Recall that our approach involves a $\gamma$ threshold for \empt detection (cf. Eq.~\eqref{eq:score-kb}).  Here, we estimate this threshold using 5-fold cross-validation, and the average $\gamma$ value is 0.8 for ED and 0.6 for MP.
The reason that edit distance performs better is that it is more robust with respect to the value of $\gamma$. In other words, a single $\gamma$ value performs well across different predicate-column heading pairs.

\begin{table}[t]
	\centering
	\caption{Value finding performance using a knowledge base. Significance of MP is tested against ED.}
	\label{table:vf-kb}
	\vspace*{-0.75\baselineskip}
	\begin{tabular}{ll@{~~}l@{~~}l@{~~}l@{~~}}
		\toprule
		Method & \multicolumn{2}{c}{\empt excluded}  & \multicolumn{2}{c}{\empt included} \\
		  & \small{NDCG@5} & NDCG@10 & NDCG@5 & NDCG@10 \\
		\midrule
		KBLookup ED & \textbf{0.5255} & 0.5308 & \textbf{0.5015} & \textbf{0.5048} \\
		KBLookup MP & 0.5222 & \textbf{0.5316}  & 0.4489$^\ddag$ & 0.4549$^\ddag$ \\
		\bottomrule
	\end{tabular}
  \vspace*{-\baselineskip}		
\end{table}

\subsubsection{Using a Table Corpus}

We consider (i) two table matching methods, InfoGather and TMatch;\footnote{Additionally, we have also considered DTS~\cite{Nguyen:2015:RSS} and the method in~\citep{DasSarma:2012:FRT} for table matching. However, both were inferior to InfoGather in terms of effectiveness. Therefore, in the interest of space we only report only on InfoGather.} 
(ii) two evidence combination strategies, top-ranked table (top) and all tables (all); and 
(iii) four heading label similarity methods, uniform (UNI), edit distance (ED), mapping probability (MP), and Label2Vec (L2V).
Table~\ref{table:vf-tc} presents all possible combinations of these. 


Our observations are as follows.
Regarding the two table matching methods (lines 1--8 vs. 9--16), we find that TMatch can outperform InfoGather by up to 18\%, with all other components being identical. Many of the differences (esp. when using all tables) are statistically significant. This shows that value finding benefits from better table matching, which is as expected.
When comparing the two evidence combination strategies (lines 1--4 vs. 5--8 and 9--12 vs. 13--16), we find the \emph{top} method to be the better overall performer.  There are a few exceptions, however, when \emph{all} delivers marginally better results, e.g., TMatch with ED, MP, or L2V, with \empt included.  
Finally, the ranking of heading label similarity methods is ED, L2V $>$ UNI $>$ MP.  That is, ED and L2V perform best, with minor differences between the two depending on the particular configuration.  Interestingly, MP does not work well, in fact, it performs even worse than not incorporating heading similarity at all (UNI).

\begin{table}[t]
	\centering
	\caption{Value finding performance using a table corpus. Highest score are boldfaced. Significance of TMatch (lines 9-16) is tested against InfoGather (lines 1-8).}
	\label{table:vf-tc}
	\vspace*{-0.75\baselineskip}
	\begin{tabular}{@{}p{1.2cm}p{0.3cm}ll@{~~}l@{~}l@{~~}l@{}}
		\toprule
		\multicolumn{3}{l}{Method} & \multicolumn{2}{c}{\empt excluded}  & \multicolumn{2}{c}{\empt included} \\
		 & & & NDCG@5 & NDCG@10 & NDCG@5 & NDCG@10 \\
		\midrule
		InfoGather & top & UNI & 0.6178 & 0.6425 & 0.5142 & 0.5370   \\
		 &  & ED & \textbf{0.6670} & \textbf{0.6854} & \textbf{0.5497} & \textbf{0.5675}     \\
		& & MP & 0.4968 & 0.5428 & 0.4474  & 0.4848   \\
		& & L2V & 0.6600 & 0.6792 & 0.5442 & 0.5634    \\
		\midrule
		InfoGather & all & UNI & 0.5992  & 0.6255 & 0.5052 & 0.5289   \\
		& & ED & 0.6445 & 0.6685 & \textbf{0.5561}  & \textbf{0.5753}   \\
		& & MP & 0.4677 & 0.5252 & 0.4348  & 0.4802   \\
		& & L2V & \textbf{0.6489} & \textbf{0.6714} & 0.5365  & 0.5576   \\
		\midrule
		TMatch & top & UNI & 0.6463$^\ddag$  & 0.6670$^\ddag$ & 0.5342$^\dag$ & 0.5524$^\ddag$   \\
		& & ED & \textbf{0.6930}$^\ddag$  & \textbf{0.7077}$^\ddag$  & 0.5626 & 0.5772 \\
		& & MP & 0.5256$^\ddag$  & 0.5790$^\ddag$  & 0.4664 & 0.5096 \\
		& & L2V & 0.6863$^\ddag$  & 0.7028$^\ddag$  & \textbf{0.5630}$^\ddag$  & \textbf{0.5791}$^\ddag$  \\
		\midrule
		TMatch & all & UNI & 0.6208$^\ddag$   & 0.6459$^\ddag$  & 0.5335$^\ddag$ & 0.5534$^\ddag$   \\
		& & ED & 0.6402  & 0.6692  & 0.5534   & 0.5753    \\
		& & MP & 0.5234 & 0.5427$^\ddag$  & 0.4788   & 0.4921    \\
		& & L2V & \textbf{0.6739}$^\ddag$  & \textbf{0.6921}$^\ddag$  & \textbf{0.5678}$^\ddag$   & \textbf{0.5851}$^\ddag$    \\

		\bottomrule
	\end{tabular}
  \vspace*{-0.5\baselineskip}	
\end{table}
%


\subsection{Feature Importance Analysis}
\label{sec:eval:anal}

\begin{figure*}[t] 
  \centering
  \vspace*{-0.75\baselineskip}
  \includegraphics[width=0.9\linewidth]{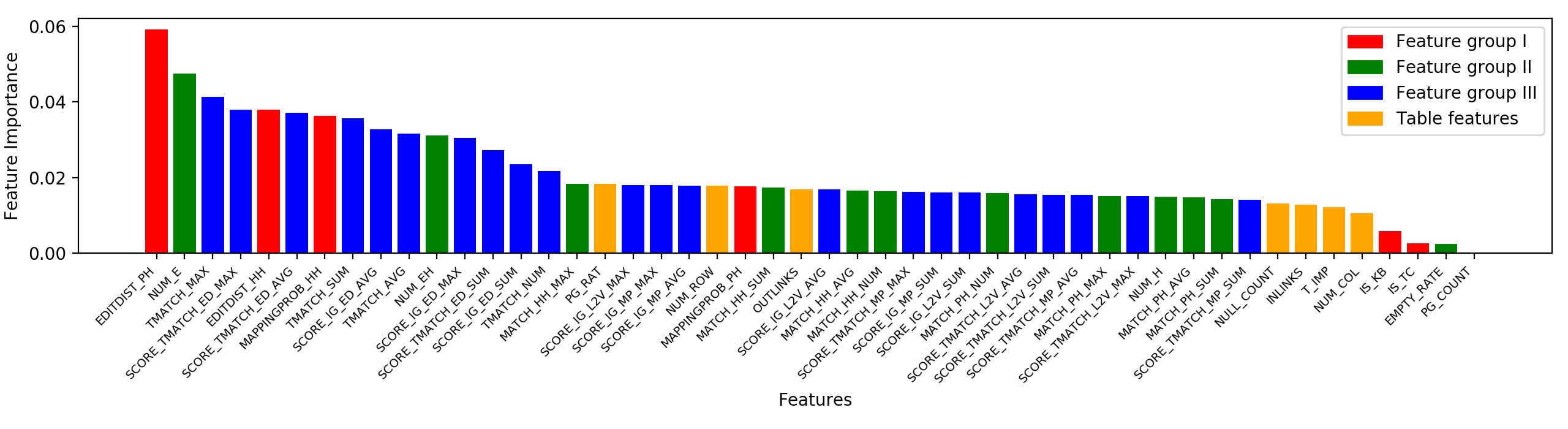} 
  \vspace*{-1.5\baselineskip}
  \caption{Feature importance measured in terms of Gini score.}
  \label{fig:feat_imp}
  \vspace*{-0.5\baselineskip}
\end{figure*}

In order to gain an understanding of which features contribute most to the effectiveness of our value ranking approach, we measure their importance in terms of Gini score.  The results are shown in Fig.~\ref{fig:feat_imp}, ordered left to right from most to least important.
Generally, features from group I and III are the most represented at the top ranks, while feature group II and table features dominate the bottom half of the ranking.

%

\subsection{Cell-level Analysis}
\label{sec:eval:examples}

So far, we have reported on aggregate statistics.  In our final experimental section, we perform an analysis on the level of individual cells.
Recall that when creating the test collection, we have concealed the original cell values from the input tables, pretending that these were missing.  In this part, we compare these original cell values (referred to as \emph{original}) with the values that were retrieved automatically by our approach (referred to as \emph{found}).  


Table~\ref{tbl:vv_stat} reports the overall statistics.  The first and second lines of this table represent the cases where the cell was originally empty and had a value in the input table, respectively.  The columns of the table correspond to how many different (valid) values were found by our approach. Below, we take a closer look at each of these cases, from top to bottom and from left to right.

\begin{table}[t]
	\centering
	\small
	\caption{Cell-level analysis, comparing the original (concealed) values in test tables against those found by our \textsc{Cell\-AutoComplete} method.}
	\label{tbl:vv_stat}
	\vspace*{-0.75\baselineskip}
		\begin{tabular}{l|*{3}{c}}
		\hline
		\diagbox{Original}{Found}
		&\makebox[3em]{\textbf{0}}&\makebox[3em]{\textbf{1}}&\makebox[3em]{\textbf{2+}}\\
		\hline
		\textbf{0} & 20 & 3 & 5 \\
		\textbf{1} & 154 & 205 & 613\\\hline
	\end{tabular}
  \vspace*{-1\baselineskip}
\end{table}

\begin{itemize}
	\item There are 20 cells, where originally the cell was empty and we also did not find a value (i.e., no difference). 
	\item In 3 cases, we found the value for a cell that was originally empty. On such example is the ``departure'' time for ``Hampton Roads'' in a table about ``Itinerary.''
	\item In 5 cases, we identified two valid values for a cell that was originally empty. For instance, for the ``type'' column of ``Polvorones'', in a table about ``Breads and pastries,'' both ``shortbread'' and ``bread'' are correct values. 
	\item There are 154 cells where the cell is originally not empty, but we could not find its value. This is the category where our method failed.  It turns out that in most of these cases, the given values exist only in the original tables (which were excluded from the corpus). 
	\item In 205 cases, both \emph{original} and \emph{found} have the same single value (i.e., no difference).
	\item For 613 cells that are originally non-empty, we found multiple valid values.  In many cases, \emph{found} includes further values in addition to the original value.  E.g., the original value is ``Republican Party (United States),'' while the found values also include ``R'' and ``Republican.'' Another example is an athlete's ``country,'' which is originally ``Brazil at the Olympics'', while the found values also include ``Brazil.''  In some cases the granularity of the values differ, e.g., the ``location'' of ``Pike'' is ``Levee Township, Pike County, Illinois'' in the original table, while values in \emph{found} also include ``Hull, Illinois'', ``Detroit, Illinois'', ``Pittsfield, Illinois'', and ``Pearl, Illinoi.''
	In other cases, there is no overlap between the values returned by \emph{original} and \emph{found}.  There are several cases where the difference is in the value formats or in the granularity.   E.g., the original table contains ``1982'' as the ``death'' date of ``Hugh John Flemming,'' while the value we returned from the knowledge base is ``1982-10-16.''  Another reason for the differences has to do with temporal mismatch, i.e., one of the sources is out-of-date.
	  Finally, there are also some genuine cases of conflicting values. E.g., the ``open date'' of ``Kannon Station'' is ``1923-07-05'' in the original table, while in DBpedia the ``opening year'' is ``1913-01-01.'' Similarly, the ``Platform'' of ``Okular'' is ``MS'' according to one Wikipedia table, while it is ``Unix-like'' in DBpedia. 
	
\end{itemize}  
\vspace*{-0.25\baselineskip}
Overall, our method finds the same as the original value in 22.5\% of the cases, misses the original value in 15.4\% of the cases, and finds either additional correct values or conflicting values in 62.1\% of the cases.  This latter category highlights the usefulness of cell auto-completion. It also suggests further potential for other applications, such as fact-checking.

\vspace*{-0.75\baselineskip}
\section{Conclusions}

We have addressed the task of auto-completing cell values, given an input relational table.  Using a knowledge base and a table corpus as sources, we have demonstrated the effectiveness of our approach on a purpose-built test collection.
While we have developed our approach with a specific application in mind, these techniques can also be utilized for other tasks, including information extraction, populating KBs from tables, and truth/fact finding.

We see several avenues for future work.  We would like to move beyond relational tables and beyond the clean and well-organized tables that can be found in Wikipedia, by considering arbitrary tables from the Web.  In addition to  structured sources, we are interested in incorporating evidence from unstructured text, e.g., web pages.  Finally, we wish to explicitly address the temporal aspects of certain entity attributes.

\bibliographystyle{ACM-Reference-Format}
\bibliography{00paper-vf}

\end{document}